\documentclass[pra, preprint, superscriptaddress, nofootinbib, showpacs]{revtex4}

\usepackage[english]{babel}
\usepackage{multirow}
\usepackage{graphicx}
\usepackage{dcolumn}
\usepackage{bm}
\usepackage{amsmath}
\usepackage{amsfonts, amssymb}
\usepackage{dsfont}
\usepackage{graphicx}
\usepackage{subfigure}
\usepackage{xcolor}
\usepackage{subfigure}

\begin{document}

\title{Wave-packet propagation based calculation of above-threshold ionization in the x-ray regime}

\author{Matthew Tilley}
 \email{matthew.tilley@desy.de, mt2911@ic.ac.uk}
\affiliation{ Center for Free-Electron Laser Science, DESY, DE-22607 Hamburg, Germany}
\affiliation{ Department of Physics, Imperial College, London SW7 2AZ, UK}

\author{Antonia Karamatskou}
 \email{antonia.karamatskou@cfel.de}
\affiliation{ Center for Free-Electron Laser Science, DESY, DE-22607 Hamburg, Germany}
\affiliation{Department of Physics, Universit\"at Hamburg, D-20355 Hamburg, Germany}

\author{Robin Santra}
 \email{robin.santra@cfel.de}
\affiliation{ Center for Free-Electron Laser Science, DESY, DE-22607 Hamburg, Germany}
\affiliation{Department of Physics, Universit\"at Hamburg, D-20355 Hamburg, Germany}


\begin{abstract}

We investigate the multi-photon process of above-threshold ionization for the light elements hydrogen, carbon, nitrogen and oxygen in the hard x-ray regime. Numerical challenges are discussed and by comparing Hartree-Fock-Slater calculations to configuration-interaction-singles results we justify the mean-field potential approach in this regime. We present a theoretical prediction of two-photon above-threshold-ionization cross sections for the mentioned elements. Moreover, we study how the importance of above-threshold ionization varies with intensity. We find that for carbon, at x-ray intensities around $10^{23}{\rm Wcm}^{-2}$, two-photon above-threshold ionization of the K-shell electrons is as probable as one-photon ionization of the L-shell electrons.

\end{abstract}


\maketitle

\section{\label{sec:level1}Introduction}

The development of x-ray free-electron lasers (XFELs) in recent years has allowed the production of ultrashort x-ray pulses at ever increasing intensities. Highly intense hard x rays are of particular interest, e.g., for the purposes of molecular imaging at atomic resolution \cite{Chapman.Nature}. Anticipating further developments in the direction of ultrashort x-ray pulses down to a few hundred attoseconds time scale ($1$~as$=10^{-18}$~s), XFELs will represent the ideal tool for single-molecule imaging via coherent x-ray scattering \cite{Nature2000,PhysRevA.83.033402}. Further interesting applications of highly intense and ultrashort pulses include the investigation of electronic dynamics in atoms and (bio-) molecules which typically take place on a time scale between attoseconds and tens of femtoseconds ($10^{-15}$~s) \cite{Nature420,RevModPhys.81.163}.

The Linac Coherent Light Source (LCLS) XFEL has been running since 2009 at the SLAC National Accelerator Laboratory in the US and was the first XFEL capable of producing hard x-rays \cite{SLAC.press}. Recently, the SACLA XFEL at the SPring-8 facility in Japan reached intensities of $10^{20}$~Wcm$^{-2}$ at photon energy of $9.9$~keV \cite{Nat.Commun.5:3539}. The European XFEL in Hamburg, Germany, is intended to produce x-rays with photon energies up to $12-100$~keV \cite{europenXFEL.design}. By focusing the pulses down to a few nanometers, XFELs can reach intensities that are orders of magnitudes greater than previously achieved. Even though the interaction probability of x-rays with matter is low \cite{xdb2001}, in this high-intensity regime it is necessary to consider the importance of nonlinear processes affecting electronic dynamics of atomic, molecular or solid-state target systems. 

The present study focuses on the nonlinear effect of above-threshold ionization (ATI). First observed in 1997 by P. Agostini {\it et al.}, ATI is a process whereby an electron absorbs more photons than are necessary for ionization \cite{PhysRevLett.42.1127}. This process has been studied extensively by now, especially in the range of infrared to visible and XUV light \cite{J.Phys.B.39.R203,Nature.432.605,PhysRevA.70.043412}; however, less work exists for x-ray ATI in the high-intensity regime \cite{PhysRevA.80.053424}, and mostly on hydrogen or hydrogen-like ions 
\cite{PhysRevA.84.033425,Cent.Eur.J.Phys,JPhysB43,PhysRevA.86.033413}. 

The purpose of this work is to examine the role and the magnitude of ATI in the x-ray regime under the high-intensity conditions that will become available soon at XFELs. To this end, photoelectron spectra are calculated in order to quantify the effect of ATI. Our method for the calculation of photoelectron spectra is based on the time-dependent configuration interaction singles (TDCIS) method \cite{PhysRevA.82.023406,PhysRevA.89.033415,XCIDprog}. The $N$-electron wave function is obtained by solving the Schr\"odinger equation numerically on a grid. In order to prevent reflections from the end of the grid and to analyze the outgoing wave packet, the so-called wave function splitting method is employed. The first-principle calculations of the TDCIS method are compared to calculations treating the atomic potential on the Hartree-Fock-Slater (HFS) level. For the purposes of molecular imaging the most important elements to consider are those commonly found in organic molecules. For this reason carbon, nitrogen, oxygen and hydrogen were chosen as the focus of this study. The two-photon ATI cross sections for these elements were calculated at the representative and commonly used hard x-ray photon energies of $8$, $10$ and $12$~keV. An investigation was also carried out to see at which intensities ATI makes an important contribution to the overall ionization probability. 

In Sec.~\ref{sec:2} a discussion of the theoretical basis of the method is presented, followed by the numerical and computational challenges faced during the investigation of photon-atom interactions in the x-ray regime. In Sec.~\ref{sec:3} both the two-photon ATI cross sections for light elements and the results of an intensity study for ATI in carbon are shown. A short summary in Sec.~\ref{sec:4} concludes this work. Atomic units are used throughout unless otherwise stated.

\section{\label{sec:2}Theory and Method}

\subsection{Theoretical Overview}

The time-dependent Schr\"odinger equation for a $N$-electron system is given by
\begin{equation}\label{Sch}
i \frac{\partial}{\partial t} |\Psi(t)\rangle= \hat H |\Psi(t)\rangle.
\end{equation}
Here $|\Psi(t)\rangle$ is the full $N$-electron wave function. The Hamiltonian $\hat{H}(t)$ is of the form
\begin{equation}\label{Ham}
\hat{H}(t)=\hat{H}_{0}+\hat{H}_{1}+\hat{\mathbf{p}} \cdot \hat{\mathbf{A}}(t),
\end{equation}
where $\hat{H}_{0}= \hat{T} + \hat{V}_{\rm nuc} + \hat{V}_{\rm mf} - E_{HF}$, containing terms for the kinetic energy $\hat{T}$, the nuclear potential $\hat{V}_{\rm nuc}$, the mean-field potential $\hat{V}_{\rm mf}$,  and the Hartree-Fock energy $E_{\rm HF}$. The part of the exact electron-electron Coulomb interaction $\hat{V}_{\rm c}$ not included within $V_{\rm mf}$ is provided by $\hat{H}_{1}=\hat{V}_{\rm c}-\hat{V}_{\rm mf}$. Under the dipole approximation $ \hat{\mathbf{p}} \cdot \hat{\mathbf{A}}(t)$ describes the light-matter interaction in the velocity form (the light is assumed to be linearly polarized).

The solution to Eq.~\eqref{Sch} is written as a superposition of the Hartree-Fock ground state $|\Phi_{0}\rangle$ and all 1-particle--1-hole excitations $|\Phi_{i}^{a}\rangle$
\begin{equation}
|\Psi(t)\rangle=\alpha_{0}(t)|\Phi_{0}\rangle+\sum_{i,a}\alpha_{i}^{a}(t)|\Phi_{i}^{a}\rangle, \label{wave function}
\end{equation}
where the index $i$ specifies an initially occupied orbital and $a$ indicates an unoccupied orbital to which the electron can be excited. The 1-particle--1-hole excitations are given by
\begin{equation}\label{1p1h}
|\Phi_{i}^{a}\rangle=\frac{1}{\sqrt{2}}\left\{ \hat{c}_{a+}^{\dagger}\hat{c}_{i+}+\hat{c}_{a-}^{\dagger}\hat{c}_{i-}\right\} |\Phi_{0}\rangle.
\end{equation}
Here, $\hat{c}^{\dagger} $ and $\hat{c}$ represent creation and annihilation operators for the corresponding orbital, respectively. By applying the time-dependent Schr\"odinger equation to Eq.~\eqref{wave function} and projecting onto the $|\Phi_{0}\rangle$ and $|\Phi_{i}^{a}\rangle$ states we obtain an equation of motion for the expansion coefficients $\alpha_{0}(t)$ and $\alpha_{i}^{a}(t)$. The TDCIS method as described here assumes closed shell atoms in the Hartree-Fock ground state, such that the total spin is zero. A detailed description of TDCIS can be found in Refs.~\cite{PhysRevA.82.023406,PhysRevA.85.023411}.

In order to both eliminate reflections and to store information about the outgoing wave packet, the method of wave-function splitting was used, which was first implemented by Tong {\it et al.} \cite{PhysRevA.74.031405}. Briefly, a splitting operator $\hat{\rm S}$ is applied to the wave function that has the shape of a smoothed-out step function
\begin{equation}
\hat{\rm S}=\left\{ 1+\exp\left[-(\hat{r}-r_{c})/\Delta\right]\right\}^{-1},
\end{equation}
where the parameter $\Delta$ controls the smoothness of the function and $r_{c}$ determines the center. The wave function is then split into $|\chi_{\rm in}(t_{0})\rangle$ and $|\chi_{\rm out}(t_{0})\rangle$ where the inner and outer parts are treated separately: 
\begin{equation}
\begin{split}
|\chi_{\rm in}(t_{0})\rangle&=(1-\hat{\rm S})|\chi_{i}(t_{0})\rangle ,\\ \\
|\chi_{\rm out}(t_{0})\rangle&=\hat{\rm S}|\chi_{i}(t_{0})\rangle.
\end{split}
\end{equation}
Here, $t_0$ is the current time step and $|\chi_{i}(t_{0})\rangle$ is the wave function for a particular ionization channel $i$. By absorbing and independently propagating the outer part analytically to long times, reflections from the end of the grid are avoided. A more detailed description of the calculation of photoelectron spectra (PES) using the TDCIS approach can be found in Ref.~\cite{PhysRevA.89.033415}. In order to compute cross sections for open shell atoms, we use here the HFS atomic potential \cite{PhysRev.81.385}. The HFS potentials for the various elements were calculated using the XATOM code \cite{Son13e}.

\subsection{Numerical Challenges}

During the investigation of the interaction of x-rays with atoms a variety of computational and numerical challenges arise. The grid on which the wave function is represented requires a large radius in order to efficiently apply the splitting method to the high-energy wave packets produced. Grid sizes of around 120 Bohr radii were found to be large enough. The number of grid points was chosen at approximately 10 points per de-Broglie wavelength in order to well represent the high-energy parts of the outgoing wave function. Increasing the number of grid points further did not influence the PES. It was found that for the x-ray photon-energy regime choosing a maximum angular momentum of higher than 3 did not significantly affect the PES.

Because of the high photon energy the propagation time step used to propagate the wave function needs to be very small. We found that using Runge Kutta to the 4th order about 20 time steps per electric field oscillation are required to prevent significant artifacts from appearing in the PES. As a consequence of the small propagation time step the splitting function had to be applied very frequently. Applying the splitting every 3 time steps is appropriate to remove artifacts due to reflections. However, this causes some practical problems as large amounts of data must be stored. 
Certain parameters are found to be mostly unimportant for convergence. Consistently with the study in Ref.~\cite{PhysRevA.89.033415}, the smoothing of the splitting function $\Delta$ has little to no effect as long as the wave function close to the nucleus is not disturbed.

\begin{figure*}[!t]

\includegraphics[width=\linewidth]{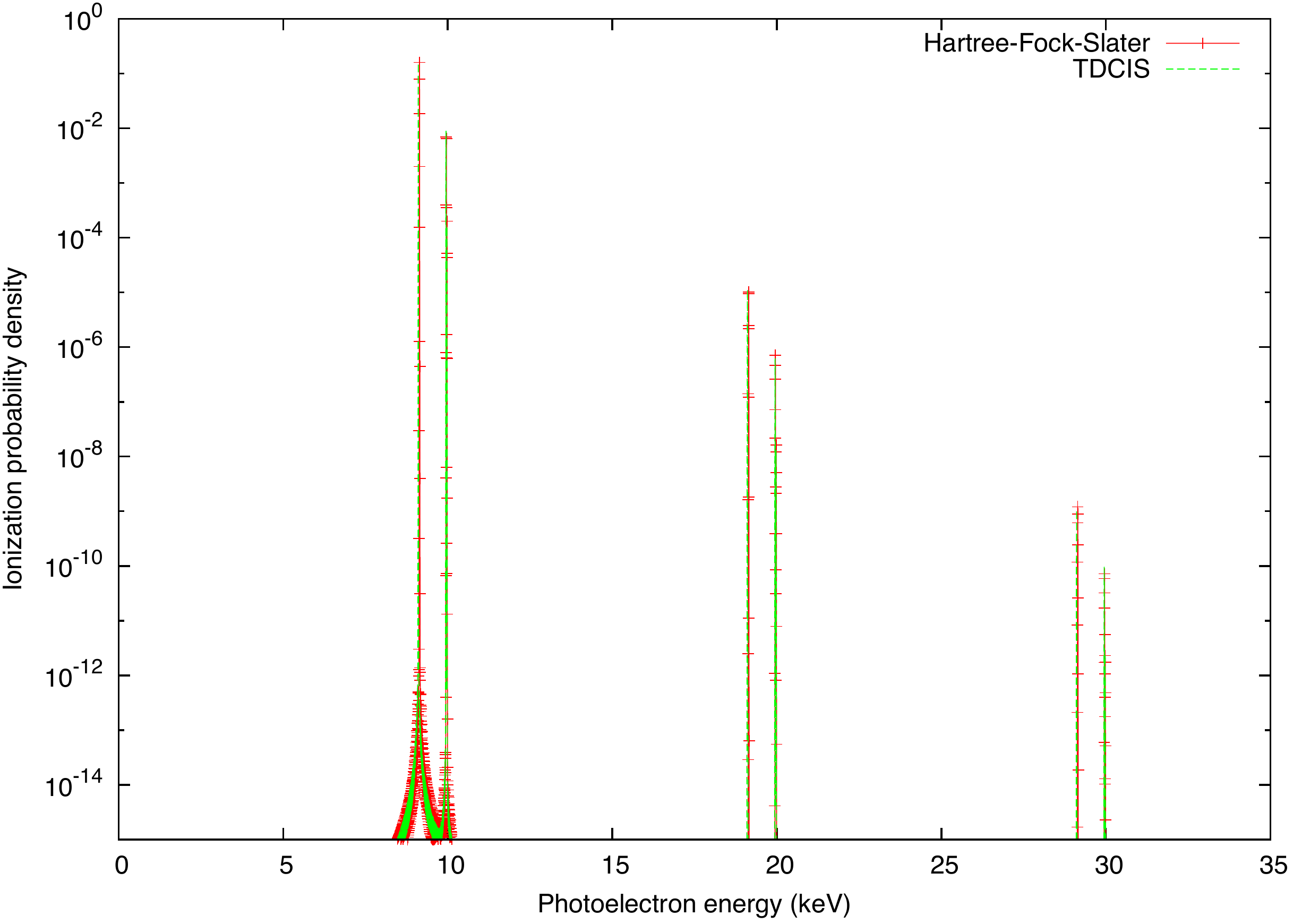}
\caption{(Color online) Photoelectron spectrum of neon for a Gaussian pulse with $10$~keV photon energy, at $10^{21}$~Wcm$^{-2}$ intensity and $0.48$~fs pulse duration, using HFS (green) and TDCIS (red). The one-photon ionization peak the first two ATI peaks are shown. Each peak consists of two subpeaks, the one at lower energy being associated with ionization from the K shell, the other one corresponding to L-shell ionization.}
\label{fig.hfs}
\end{figure*}

Obviously, using the HFS potential significantly reduces the computational time as it removes the necessity to calculate the large number of Coulomb matrix elements, i.e., the exact Coulomb interaction between the electrons. In order to find out if the HFS approach is valid in the x-ray photon-energy range, neon was studied with both methods before applying it to the elements carbon, nitrogen and oxygen, which are more relevant for chemistry and biology than nobel-gas atoms. In Fig.~\ref{fig.hfs} we present the photoelectron spectrum of neon for a Gaussian pulse of $10$~keV photon energy and $0.48$~fs pulse duration at an intensity of $10^{21}$~Wcm$^{-2}$ using the HFS potential (green) and the full TDCIS method (red). Except for the slight upward energy shift for HFS the shape and height of the peaks are the same. The noise levels and artifacts present in the spectrum are also unaffected by the method used. The comparison suggests that in the photon energy regime of hard x-rays the HFS approach is well justified for open shell atoms with a similar $Z$ value, namely carbon, nitrogen, and oxygen. Our calculations for neon show that electron correlation effects play a minor role for one- and two-photon absorption in this regime. 

\section{\label{sec:3}Results and Discussion}
We calculated two-photon ATI cross sections for hydrogen and compared them with previous work \cite{PhysRevA.84.033425}. They are found to be in very close agreement. Using $10^{20}$~Wcm$^{-2}$, a relative difference of $2$\% was found at the photon energy of $10$~keV and $7$\% at $8$~keV. With a slightly larger intensity of $3.5\times10^{20}$~Wcm$^{-2}$ the cross section at $5$~keV was found to be $2.02\times10^{-65}$~cm$^{4}$s, which amounts to a difference of $1$\% compared to Ref.~\cite{PhysRevA.84.033425}. Since the method employed here relies on the numerically exact solution of the Schr\"odinger equation \eqref{Sch}, no difficulties arise with sums over intermediate states, which appear in a perturbative treatment \cite{PhysRevA.80.053424}. Reference \cite{PhysRevA.84.033425} demonstrates that the $A^2$ interaction, which is not included here, may be neglected in the photon-energy range of current interest. As already indicated in Eq.~\eqref{Ham} by the exclusive time dependence of the vector potential all calculations were performed under the assumption of the dipole approximation. However, under the conditions of short-wavelength x-rays this assumption may no longer be valid. Indeed, Zhou and Chu indicate that nondipole effects significantly change the photoelectron angular distribution and that the nondipole ATI spectra are enhanced in the high photon-energy regime \cite{PhysRevA.87.023407}. On the other hand, it was shown for the x-ray regime that when including all multipoles the {\em total} two-photon ATI cross section differs less than an order of magnitude from the cross section calculated in dipole approximation for sufficiently small nuclear charge \cite{PhysRevA.86.033413}. Therefore, although our values may underestimate the two-photon ATI cross section, we expect this underestimation in the integrated spectrum to be much smaller than an order of magnitude (a factor of around $2-3$) \cite{PhysRevA.85.013423}.

The two-photon ATI cross sections were found in the perturbative limit, i.e., by assuring that the ionization probability is low enough to not deplete the ground state. In the perturbative limit the ionization probability due to 2-photon absorption $\rho^{(2)}$ is given by $\sigma^{(2)}F^{(2)}=\sigma^{(2)}\int J^2 {\rm d} t$, where $\sigma^{(2)}$ is the two-photon cross section and $J$ is the photon flux in cm$^{-2}$s$^{-1}$; $F^{(2)}$ is the fluence for two-photon absorption. Assuming a Gaussian pulse the electric field has the form
\begin{equation}
E(t)=E_0\cos(\omega t)e^{-2\ln 2\ t^2/\tau^2},
\end{equation}
where $E_0$ is the peak electric field, $\omega$ is the central field frequency, and $\tau$ is the pulse duration. Then, the fluence $F^{(2)}$ is given by
\begin{equation}\label{eq:flu}
F^{(2)}=\frac{E_{0}^{4}}{\omega^{2}}\sqrt{\frac{\pi}{8\ln2}}\left(\frac{c}{8\pi}\right)^{2}\tau.
\end{equation}

The two-photon ATI cross sections for hydrogen, carbon, nitrogen, oxygen, and neon at the hard x-ray energies of $8$, $10$, and $12$~keV are presented in Table~\ref{tab}. We find the expected increase in the two-photon ATI cross section for higher $Z$ values and a drop with larger photon energy. As mentioned previously, all cross sections were found under the dipole approximation, which is expected to slightly underestimate the cross sections. The two-photon ATI cross section of beryllium was also calculated at $10$~keV photon energy to be $1.25\times10^{-63}$~cm$^{4}$s.

\begin{table}[!t]
\tiny
\caption{Two-photon ATI cross sections for the light elements (E is the photon energy). All were calculated from integrating the corresponding photoelectron peaks at $10^{20}$~Wcm$^{-2}$ intensity and $0.12$~fs of pulse duration.\\}
\begin{tabular}{|c|c|c|c|c|c|}
\hline 
\multirow{2}{*}{E (keV)} & \multicolumn{5}{c|}{Two-photon cross sections (cm$^{4}$s)}\tabularnewline
\cline{2-6} 
 & Hydrogen & Carbon & Nitrogen & Oxygen & Neon\tabularnewline
\hline 
\hline 
8   & $1.44\times10^{-66}$ & $1.64\times10^{-62}$ & $3.21\times10^{-62}$ & $5.62\times10^{-62}$ & $1.65\times10^{-61}$\tabularnewline
\hline 
10 & $4.69\times10^{-67}$ & $4.61\times10^{-63}$ & $9.23\times10^{-63}$ & $1.70\times10^{-62}$ & $3.71\times10^{-62}$\tabularnewline
\hline 
12 & $1.72\times10^{-67}$ & $1.79\times10^{-63}$ & $3.82\times10^{-63}$ & $6.94\times10^{-63}$ & $1.95\times10^{-62}$\tabularnewline
\hline 
\end{tabular}\label{tab}
\end{table}

\begin{figure*}[!t]
\includegraphics[width=\linewidth]{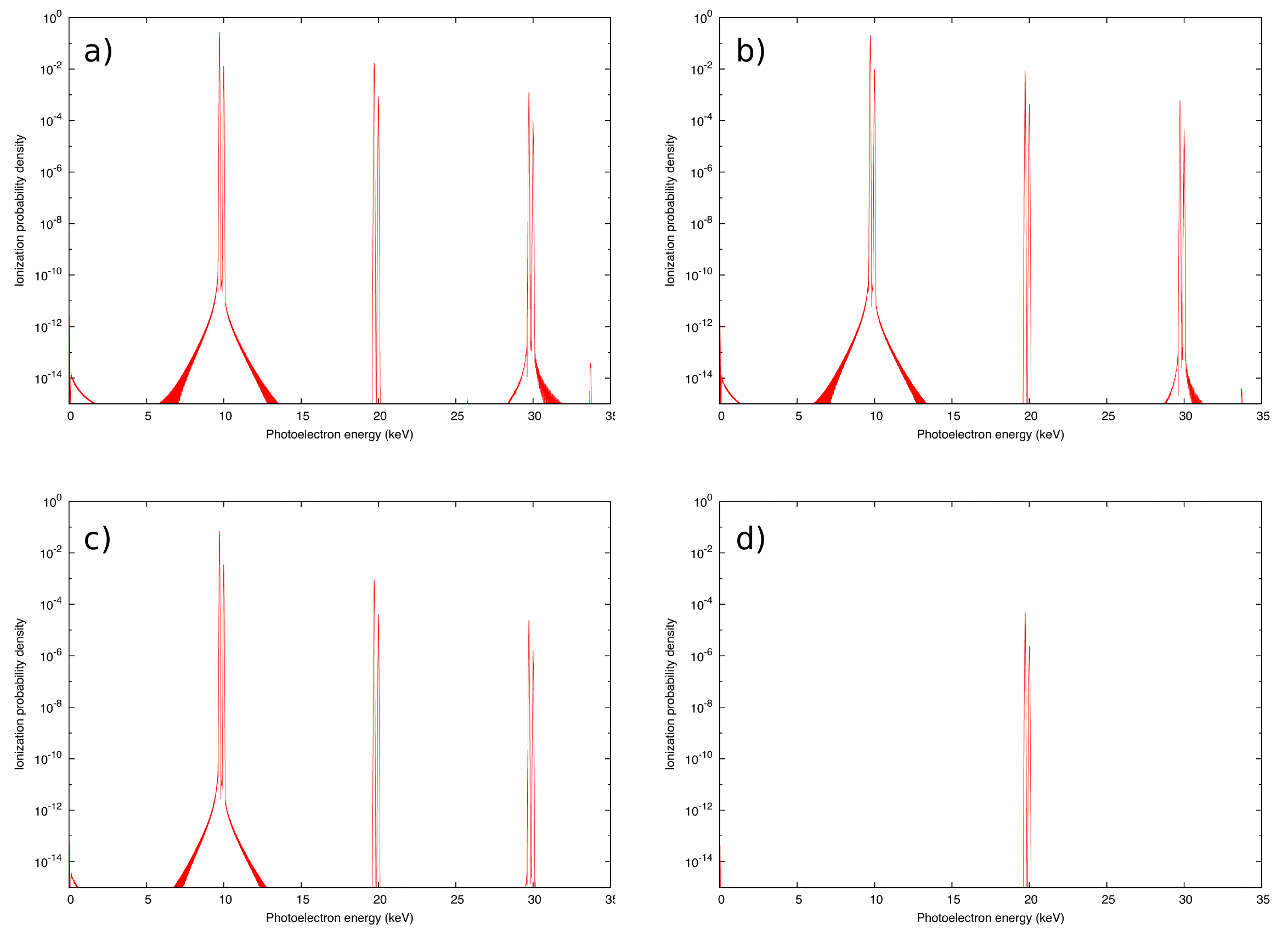}
\caption{(Color online) Photoelectron spectra of carbon for different angles: a) $0$, b) $\pi/6$, c) $\pi/3$, d) $\pi/2$. The spectra are calculated for a pulse centered at $10$~keV photon energy, at an intensity of $10^{24}$~Wcm$^{-2}$ and $0.12$~fs pulse duration. Note that the height of the K-shell ATI peak is greater than or comparable to the one-photon L-shell ionization peak.}
\label{fig.car}
\end{figure*}

\begin{figure}[!ht] 
\includegraphics[width=\linewidth]{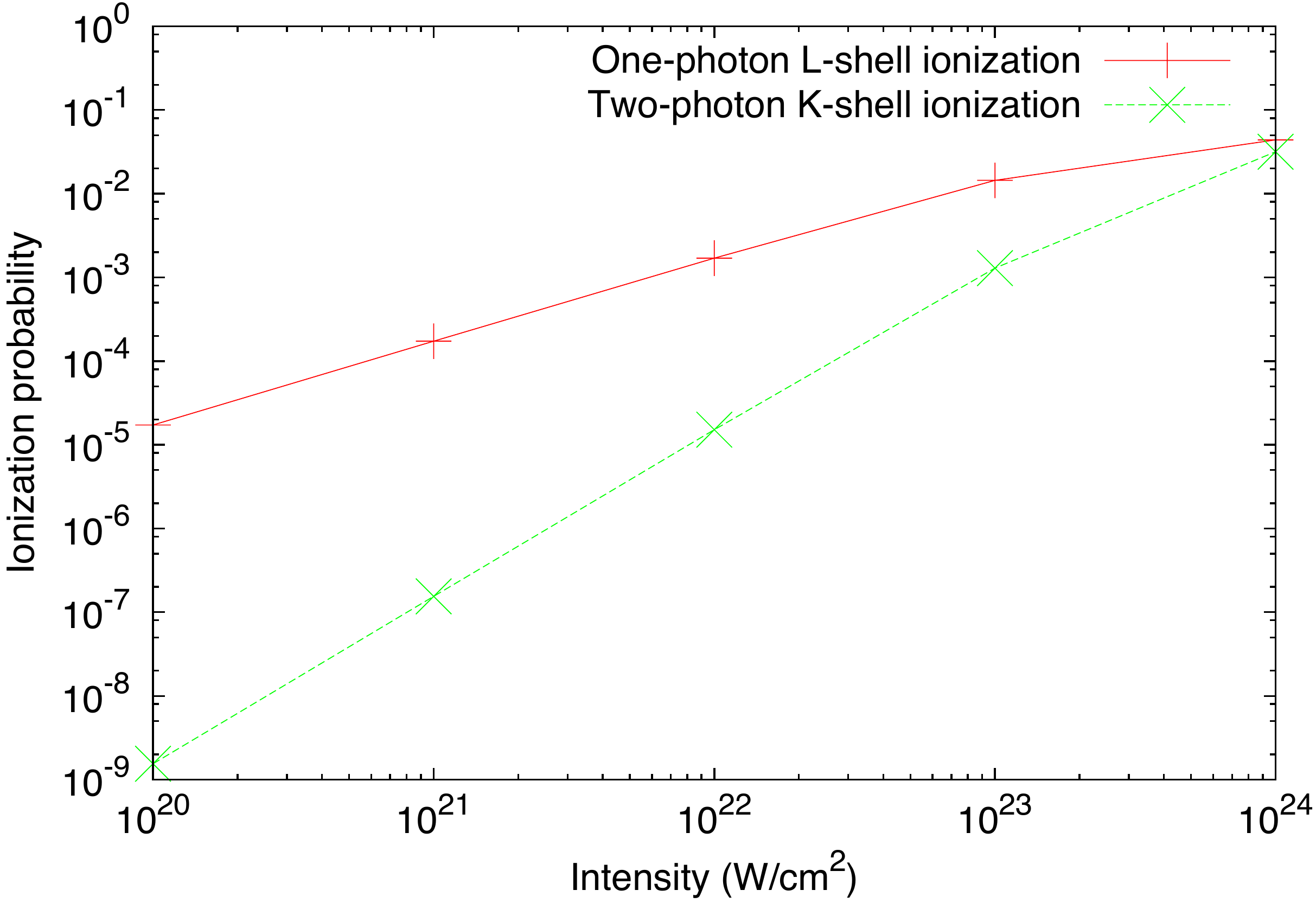}
\caption{(Color online) Depopulation at different intensities due to two-photon ATI against one-photon valence electron ionization for carbon, at $10$~keV photon energy and $0.48$~fs pulse duration. The ionization probability was found by integrating the corresponding PES peaks.}
\label{fig.int}
\end{figure}

\begin{figure}[!ht] 
\includegraphics[width=\linewidth]{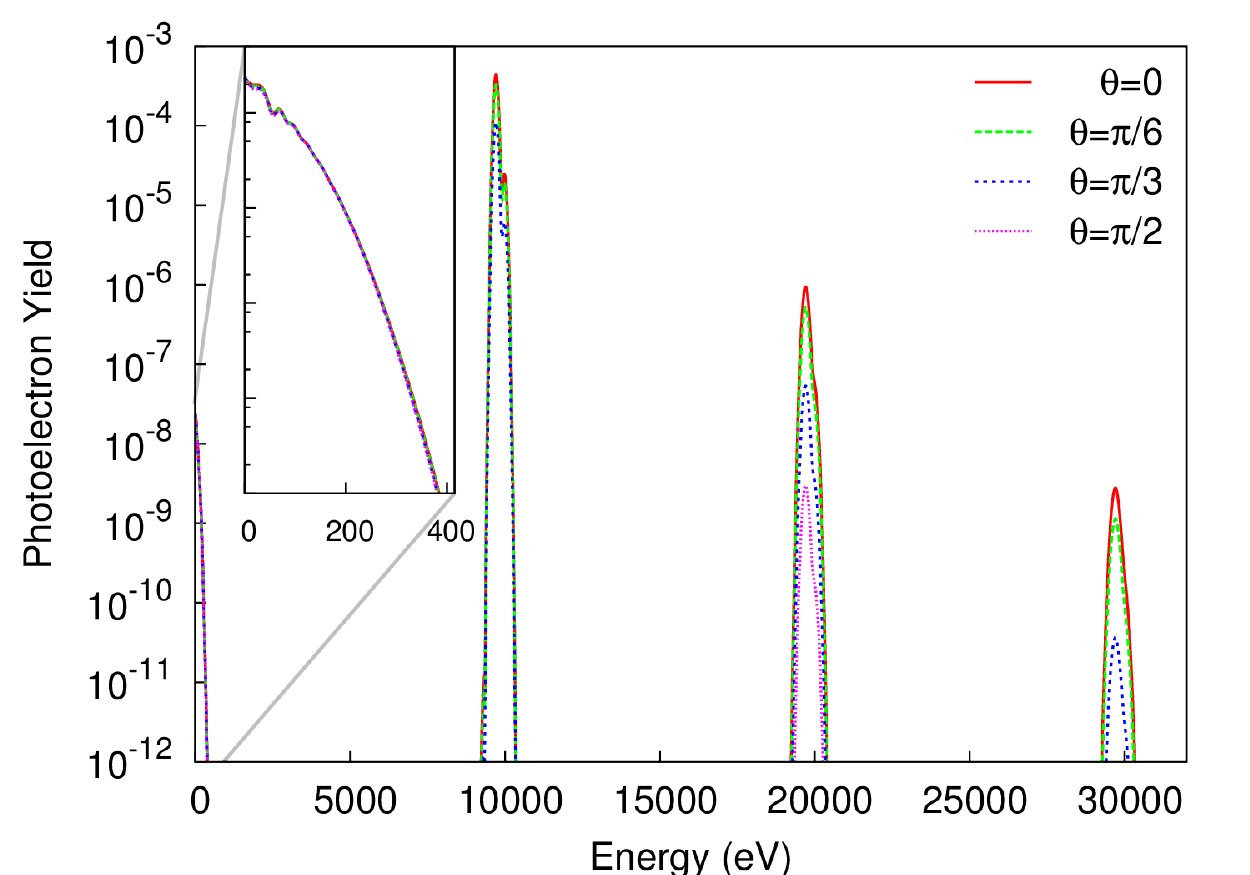}
\caption{(Color online) Photoelectron yield for carbon, shown for a pulse of $12$~as duration and $3.5\cdot10^{22}$~Wcm$^{-2}$ intensity with a photon energy of $10$~keV after a sufficiently long propagation time. The slow-electron peak (magnified in the inset)  as well as the $1$-photon peak and the first two ATI peaks are shown for $4$ different angles with respect to the polarization direction. The height of the slow-electron peak is comparable to the first ATI peak in the directions of $\pi/3$ and $\pi/2$.}
\label{sep}
\end{figure}

Because imaging of organic molecules is of particular interest we perform an intensity study on carbon with an incoming photon energy of $10$~keV in order to find the regime in which the ionization due to two-photon ATI is of the same order as that for one-photon ionization. 
As seen in Fig.~\ref{fig.car}, at an intensity of $10^{24}$~Wcm$^{-2}$ the depopulation due to K-shell ATI in the direction $ \theta=0$ becomes higher than valence one-photon ionization. The PES in the directions $\pi/6$, $\pi/3$, and $\pi/2$ are also shown (for better visualization only $15$ orders of magnitude are shown). Note, that a small peak around zero kinetic energy is observed \cite{toyota1,toyota2}. However, the height of the peak might be underestimated, because due to our splitting approach the propagation time must be sufficiently large in order to detect all electrons of interest in the splitting region. The results shown in Fig.~\ref{fig.car} were produced in a calculation where the propagation time was approximately $20$ times shorter than necessary to collect all electrons with a kinetic energy on the scale of $1$~eV. In order to elucidate this slow-electron peak further, a new calculation was performed, now involving a pulse at $10$~keV photon energy, of $12$~as duration and $3.5\cdot10^{22}$~Wcm$^{-2}$ intensity in order to be able to propagate long enough and to observe the slow-electron peak in its full height. The results are presented in Fig.~\ref{sep}. The emergence of the peak can be attributed to the bandwidth of the pulse which spans the binding energy of the valence electrons: after the absorption of one photon by the valence shell of the atom the emission of a photon can occur and, thereby, slow electrons are produced. 

In Fig.~\ref{fig.int} the ionization probability of carbon is shown as a function of intensity for the case of one-photon ionization out of the L shell together with two-photon ionization out of the K shell. We see the characteristic quadratic behavior of the ATI peak as a function of the intensity and the linear behavior in the one-photon valence ionization probability. Saturation effects do not play a role until the intensity range near $10^{24}$~Wcm$^{-2}$ is reached. In fact, one can see that at an intensity between $10^{23}-10^{24}$~Wcm$^{-2}$ the fully angle- and energy-integrated K-shell ATI peak, i.e., the ionization probability due to ATI out of the K shell, is comparable to the probability to ionize with one photon out of the valence shells.

\section{\label{sec:4}Conclusion}
We have presented a prediction for the two-photon ATI cross sections of the light elements carbon, nitrogen, and oxygen at hard x-ray energies common for current experiments at XFELs. At intensities that one may reach for future hard x-ray experiments, scientists should consider how ATI will affect their results. We conclude that ATI remains a negligible fraction of ionization for intensities at the most recent XFEL experiments with hard x-rays. However, we predict that with photon energies at around $10$~keV, when entering the regime of $10^{23}$~Wcm$^{-2}$ and above, the ionization probability of the core electrons by ATI approaches the same order of magnitude as valence stripping by one-photon ionization for elements with a similar nuclear charge $Z$ as carbon. It is likely that the neglected nondipole effects enhance the ATI spectrum by a factor smaller than an order of magnitude. Therefore, we can present our results as a lower limit on the importance of ATI and suggest that ATI be taken into account when entering this high-intensity regime. In particular, we hope that the data presented can be a guide for future experiments investigating imaging and nonlinear x-ray optics.

\bibliographystyle{apsrev}
\bibliography{bibl}

\end{document}